\documentclass[conference]{IEEEtran}
\IEEEoverridecommandlockouts
\usepackage{amsmath,amssymb,amsfonts}
\usepackage{algorithmic}
\usepackage{graphicx}
\usepackage{textcomp}
\usepackage{xcolor}
\usepackage{amsmath,graphicx}
\usepackage[nolist]{acronym} 

\usepackage{algorithm}
\usepackage{array}
\usepackage[caption=false,font=normalsize,labelfont=sf,textfont=sf]{subfig}
\usepackage{textcomp}
\usepackage{stfloats}
\usepackage{url}
\usepackage{verbatim}
\usepackage[sort]{cite}
\usepackage{tabularray}
\usepackage{tikz}
\usetikzlibrary{positioning}
\usepackage{adjustbox}
\usepackage{enumitem}   
\usepackage{pbalance}

\usepackage{booktabs}
\usepackage{hhline}
\usepackage{multirow}
\usepackage{makecell}
\usepackage{diagbox}
\usepackage{svg}
\usepackage{siunitx}
\usepackage[hidelinks]{hyperref} 

\def\BibTeX{{\rm B\kern-.05em{\sc i\kern-.025em b}\kern-.08em
    T\kern-.1667em\lower.7ex\hbox{E}\kern-.125emX}}
\begin{document}

\title{Neural Directional Filtering with Configurable \\ Directivity Pattern at Inference}

\author{\IEEEauthorblockN{ Weilong Huang$^{1}$ \qquad Srikanth Raj Chetupalli$^{2}$ \qquad Emanu{\"e}l A. P. Habets$^{1}$ }
\IEEEauthorblockA{$^{1}$\textit{International Audio Laboratories Erlangen$^{\star}$, Erlangen, Germany. 
\thanks{$^{\star}$A joint institution of Fraunhofer IIS and Friedrich-Alexander-Universität Erlangen-Nürnberg (FAU), Germany.} } \\ $^{2}$\textit{Indian Institute of Technology Bombay, Mumbai, India.}}
}

\maketitle

\begin{acronym}[DNN]
	\acro{DNN}[DNN]{deep neural network}
\end{acronym}
\begin{acronym}[iSTFT]
	
	\acro{sdri}[$\Delta$SDR]{improvement in \ac{SDR} over the unprocessed signal}
	\acro{DMA}[DMA]{differential microphone array}
	\acro{DNN}[DNN]{deep neural network}
	\acro{DOA}[DOA]{direction-of-arrival}
	\acrodefplural{DOA}[DOAs]{directions-of-arrival}
	
	\acro{iSTFT}[iSTFT]{inverse short-time Fourier transform}
	\acro{CDMA}[CDMA]{circular \ac{DMA}}

	\acro{LDMA}[LDMA]{linear \ac{DMA}}
        \acro{LS}{least-squares}
	\acro{LSTM}[LSTM]{long short-term memory}
        \acro{BiLSTM}[BiLSTM]{bidirectional long short-term memory}
        \acro{UniLSTM}[UniLSTM]{unidirectional long short-term memory}
        \acro{WNG}[WNG]{white noise gain}
	\acro{RIRs}[RIRs]{room impulse responses}
	\acro{RIR}[RIR]{room impulse response}
        \acro{RTF}[RTF]{room transfer function}
        \acro{RTFs}[RTFs]{room transfer functions}        
	\acro{DPIR}[DPIR]{direct-path impulse response}
        \acro{MVDR}[MVDR]{minimum variance distortionless response}
        \acro{LCMV}[LCMV]{linear-constraint minimum-variance}
        \acro{PMWF}[PMWF]{parametric multichannel wiener filter}
        \acro{GSC}[GSC]{Generalized sidelobe canceller}
        \acro{FT-JNF}[FT-JNF]{joint spatial and temporal-spectral non-linear filtering}  
        \acro{JNF}[JNF]{joint non-linear filtering }        
        \acro{SSF}[SSF]{spatially selective
deep non-linear filter } 
	\acro{SDR}[SDR]{signal-to-distortion ratio}
	\acro{SNR}[SNR]{signal-to-noise ratio}
	\acro{STFT}[STFT]{short-time Fourier transform}
         \acro{MAE}[MAE]{mean absolute error}	
	\acro{TF}[TF]{time-frequency}
	\acro{tsdr}[SA-$\varepsilon$-tSDR]{source-aggregated and regularized thresholded \ac{SDR}}
      \acro{STOI}[STOI]{short term objective intelligibility}
    \acro{PESQ}[PESQ]{perceptual evaluation of speech quality}
	
	\acro{UCA}[UCA]{uniform circular array}
        \acro{NDF}[NDF]{neural directional filtering} 
        \acro{UNDF}[UNDF]{neural directional filtering with user-defined directivity patterns} 

        \acro{WNG}[WNG]{white noise gain}
        \acro{DF}[DF]{directivity factor}
        \acro{DI}[DI]{directivity index}
        \acro{HRTF}[HRTF]{head-related transfer function}
        \acro{ILD}[ILD]{interaural level difference}
        \acro{FiLM}[FiLM]{feature-wise linear modulation}

    \acro{VDM}[VDM]{virtual directional microphone}
\end{acronym}

\begin{abstract}
Spatial filtering with a desired directivity pattern is advantageous for many audio applications. In this work, we propose neural directional filtering with user-defined directivity patterns (UNDF), enabling spatial filtering with directivity patterns that users can configure during inference. To achieve this, we propose a DNN architecture that integrates feature-wise linear modulation (FiLM),  allowing user-defined patterns to serve as conditioning inputs. Through analysis, we demonstrate that the FiLM-based architecture enables the UNDF to generalize to unseen user-defined patterns during inference with higher directivities, scaling variations, and different steering directions. Furthermore, we progressively refine training strategies to enhance pattern approximation and enable UNDF to approximate irregular shapes. Lastly, experimental comparisons show that UNDF outperforms conventional methods.
\end{abstract}


\section{Introduction}
\label{sec:intro}
Spatial filtering with a desired directivity pattern is beneficial to many audio applications. Traditional fixed beamforming \cite{elko2000superdirectional,brandstein2001microphone,benesty2018fixed} can achieve spatial filtering with a predefined directivity pattern by linearly filtering microphone array signals, but it requires a large number of microphones and a large array aperture not only to ensure adequate performance, as measured by \ac{WNG} and \ac{DF}, but also to approximate the predefined directivity pattern accurately. More importantly, any change of the directivity pattern requires recomputing the spatial filter. Alternatively, parametric filtering methods \cite{kallinger2009spatial, thiergart2013geometry, thiergart2013informed, thiergart2014informed, 7038281} utilize instantaneous parametric estimates, such as \ac{DOA}, to compute a filter based on any given directivity pattern. However, these methods rely heavily on accurate \ac{DOA} estimation, which can be challenging, particularly in multi-source environments containing non-speech signals \cite{thiergart2012sound}.

In recent years, many \ac{DNN}-based approaches have been proposed for extracting sound sources from a predefined \cite{rezero, location_guided2, location_guided6} or user-defined angular \cite{deep_zoom, wen2025neural} region. Notable among these methods are the \ac{JNF}-based methods \cite{tesch_insights, tesch2023multi}, which utilize \ac{DNN}-estimated masks to extract a single target speaker from an angular region of interest. 
Notably, source separation or extraction is the primary focus in all these approaches, and the directivity pattern is not explicitly controlled.

Recently, \ac{NDF} \cite{ndf_iwaenc} has been proposed to achieve spatial filtering with a desired directivity pattern for a compact array and extended to steerable \ac{NDF} in \cite{steerable_NDF_FA_2025}. Here, we aim to generate a spatial recording of the acoustic scene based on a specified directivity pattern, rather than extracting a single sound source or reducing noise. In  \cite{ndf_iwaenc, steerable_NDF_FA_2025}, we have shown that the architecture used by the aforementioned \ac{JNF}-based methods\cite{tesch_insights, tesch2023multi} can also be used for this task. However, the \ac{NDF} model learns one fixed directivity pattern that can only be steered in different directions during inference.

In this paper, we propose \ac{UNDF}, which enables spatial filtering with directivity patterns that the user can configure during inference. To achieve this, the user-defined pattern is sampled and fed as conditioning information to the \ac{DNN}. For this, we explore two mechanisms: one inspired by the conditioning approach from \cite{tesch2023multi} and the other using a FiLM approach \cite{perez2018film}. We demonstrate that the FiLM-based approach enables \ac{UNDF} to generalize to unseen user-defined patterns with higher directivities, scaling variations, and different steering directions. Additionally, we enhance pattern approximation by progressively refining training strategies, enabling \ac{UNDF} to approximate irregular shapes.

\section{Problem Formulation}
\label{sec:format}
We consider an acoustic scene in an anechoic environment, where $N$~sound sources are recorded by a compact array equipped with $Q$~omnidirectional microphones. Let $Y_q[f,t]$ represent the mixture signal at the $q$-th microphone in the \ac{STFT} domain, where $f$ and $t$ denote the frequency and time indices, respectively. The mixture signal can be decomposed as
\vspace*{-0.15cm}
\begin{equation} \label{eqn:mic_sig}
    Y_q[f,t] = \sum_{n=1}^{N} X_{q,n}[f,t] + V_q[f,t],~q\in\{1,2,\ldots,Q\},
\end{equation}
where $ X_{q,n}[f,t] $ represents the signal component due to the $n$-th source at the $q$-th microphone, $ V_q[f,t] $ represents the sensor noise at the $q$-th microphone.

The objective of the \ac{UNDF} task is to capture the acoustic scene based on a user-defined directivity pattern. A directivity pattern represents the directional sensitivity of a microphone array or a directional microphone \cite{elko2000superdirectional, eargle2012microphone}, showing the spatial response to sound from different incident angles. This paper focuses on the scenario where all sound sources are located in the $x$-$y$ plane, meaning that all incident sounds have zero elevation angles. In the following, we assume the pattern to be frequency-invariant. As a result, each pattern can be represented as a vector applicable to all frequency bands. The user-defined pattern at time $t$ is given by $\Lambda_{t}(\theta)$, where $\theta$ is the azimuth angle of the incident sound. Finally, the target signal of the \ac{UNDF} task is expressed as
\vspace*{-0.15cm}
\begin{equation}\label{eqn:vdm_sig_dir}
        Z[f,t] = \sum_{n=1}^{N} \Lambda_{t}(\theta_n)  \, X_{1,n}[f,t] ,
\end{equation}
where the $\theta_n$ is the direction of arrival for the $n$-th source, and $q=1$ is used as the reference microphone. In this work, we propose a \ac{DNN}-based method to estimate the target signal using the microphone array signals.

\section{Proposed method}
\label{sec:pagestyle}

\subsection{DNN architecture and loss function}
\begin{figure}[t!] 
\centering	\includegraphics[width=0.99\linewidth]{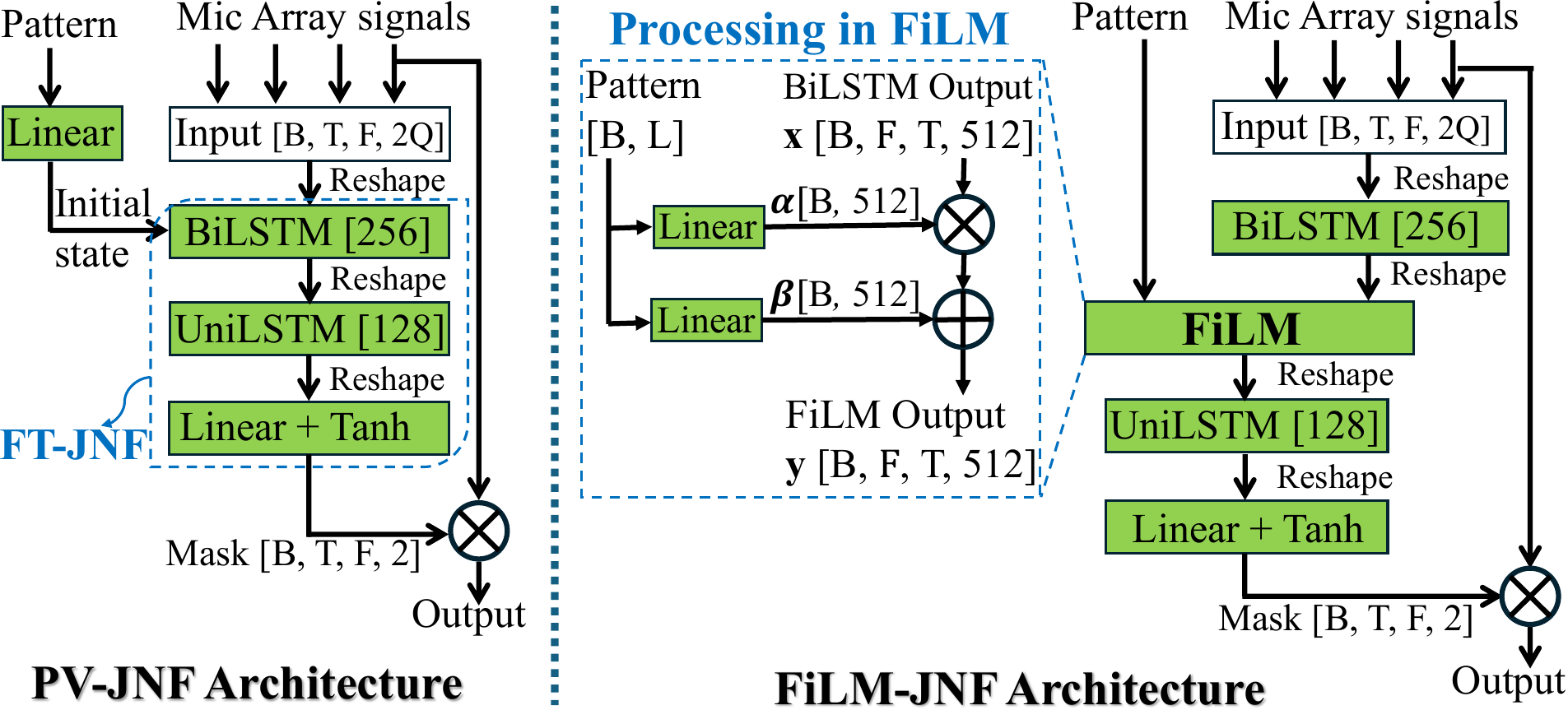}
	\caption{Proposed DNN Architectures. Left: PV-JNF, incorporating a pattern vector as the conditioning input in the \ac{FT-JNF} \cite{tesch_insights} architecture. Right: FiLM-JNF architecture, integrating FiLM \cite{perez2018film} conditioning into the \ac{FT-JNF} architecture.\vspace{-1em}}
	\label{fig: NN_arch}
\end{figure}

We propose two \ac{DNN} architectures (see Figure~\ref{fig: NN_arch}), which are adaptations of the \ac{FT-JNF} \cite{tesch_insights} for the \ac{UNDF} task. For both architectures, the real and imaginary parts of the $Q$ microphone signals in the \ac{STFT} domain are stacked along the channel dimension, resulting in an input of size $[B, T, F, 2Q]$, where $T$ represents the number of time frames, $F$ denotes the number of frequency bins, and $B$ indicates the batch size. This input is reshaped to $[B \times T, F, 2Q]$, before a \ac{BiLSTM} is applied with the \emph{frequency} axis $F$ as the recurrent sequence dimension. In addition to the microphone signals, a user-defined directivity pattern is also fed as a conditioning input. The pattern is sampled uniformly at $L$ discrete angles over the range [$0$, $2\pi$], resulting in a pattern input vector of size [B, L].  

The proposed PV-JNF architecture (left-hand side in Figure~\ref{fig: NN_arch}) is developed by modifying the \ac{SSF} framework \cite{tesch2023multi}. Original \ac{SSF} consisted of the \ac{FT-JNF} \cite{tesch_insights} and a one-hot encoding-based conditioning module for a steering angle input. In the PV-JNF, we replace the one-hot encoding with the user-defined sampled pattern vector as the input to the linear layer. The output of this linear layer serves as the initial state for the \ac{BiLSTM} layer at each time step. 

The proposed FiLM-JNF architecture (right-hand side in Figure~\ref{fig: NN_arch}) adds a dedicated condition layer between the \ac{BiLSTM} layer and the \ac{UniLSTM} to the \ac{FT-JNF} architecture \cite{tesch_insights}. Here, the condition is implemented using the \ac{FiLM} \cite{perez2018film} mechanism. The \ac{FiLM} layer computes per-feature affine parameters $\boldsymbol{\alpha}$ and $\boldsymbol{\beta}$ (with dimensions of $[B, 512]$) through two distinct linear layers from the user-defined pattern vector. Then, it applies element-wise modulation $\bf{y}=\boldsymbol{\alpha} \odot \bf{x} + \boldsymbol{\beta} $, shared across time and frequency, where $\bf{y}$ is the output of \ac{FiLM} layer, with the same dimension as $\bf{x}$. The output of the \ac{FiLM} layer is reshaped to $[B \times F, T, 512]$ and then fed to the \ac{UniLSTM} layer. Details of the remaining `Reshape' operations in Figure~\ref{fig: NN_arch} can be found in \cite{tesch2023multi, steerable_NDF_FA_2025}.

Both PV-JNF and FiLM-JNF architectures compute a complex-valued single-channel mask, denoted $\mathcal{M}[f,t]$. The estimated signal is then computed by applying this mask to the reference microphone signal, i.e., $\widehat{Z}[f,t] = \mathcal{M}[f,t] Y_{1}[f,t]$. Subsequently, we adopt a batch-aggregated normalized $\mathcal{L}_{\mathrm{1}}$-loss function for training, given by
\vspace*{-0.15cm}
\begin{equation}\label{eqn:loss_func}
\mathcal{L}_{\mathrm{1}} =  \frac{ \sum_{b=1}^B \left\| \mathbf{z}^{b} -  \hat{\mathbf{z}}^{b}   \right\|_{1}}{ \sum_{b=1}^B \left\| \mathbf{z}^{b} \right\|_{1} + \epsilon},
\end{equation}
where $\epsilon$ is a small constant value, and the time-domain signals $\hat{\mathbf{z}}$ and $\mathbf{z}$ correspond to \ac{STFT} representations $\widehat{Z}$ and ${Z}$, respectively. 

\subsection{Training strategy}
The directivity pattern of a $J$th-order \ac{DMA} can be defined as $\Lambda(\theta) =\sum_{j=0}^{J} \left(a_{j} \right.$ $\left. \times \cos^{j}(\theta - \theta_\mathrm{s})\right)$, where $a_{j}$, $j \in \{0, 1, \ldots, J \}$ are real-valued coefficients that determine the pattern's shape \cite{elko2000superdirectional}. Since \ac{NDF} approximates the shape of the mainlobe better than the sidelobes \cite{ndf_iwaenc}, and given that in sound capture scenarios users are primarily concerned with mainlobe control for the patterns while sidelobes and negative polarities are often disregarded, we consider a simplified \ac{DMA} pattern given by 
\vspace*{-0.15cm}
\begin{equation}\label{eqn:dma_new}
\Lambda(\theta) = \lvert \mu +  (1-\mu) \cos(\theta - \theta_\mathrm{s}) \rvert^{J},
\end{equation}
where $\mu$ is a real-valued number with $\mu \in [0, 1]$, which determines the null position. The patterns described by \eqref{eqn:dma_new} fulfill $\Lambda(\theta) \in [0, 1]$ (no negative polarity) and focus on the shape of the mainlobe, which is controlled by the order $J$ and $\mu$. Another widely used pattern, especially in the context of neural spatial filtering, is the rectangular pattern, defined as $\Lambda(\theta) = 1$ if $\theta \in [\theta_{\mathrm{start}}, \theta_{\mathrm{end}}]$, otherwise $0$.

The microphone array is positioned in a simulated anechoic room. For a source-array setup, we randomly select $N$ \acp{DOA} for $N$ sources. The sources and array are co-planar with a fixed source-array distance of $d$. We then simulate direct-path transfer functions for all $Q$ microphones and $N$ speakers using the RIR generator \cite{RIRGenerator} with a reflection order of zero. Subsequently, we obtain microphone signals. During training, a gain is assigned for each source signal according to the pattern vector, and the target signal is generated during training using \eqref{eqn:vdm_sig_dir}. This target signal, along with its corresponding pattern and microphone signals, is input as a training sample to the model. Each source-array setup has $P$ patterns for training, meaning there are $P$ training samples per source-array setup. The $P$ patterns are composed according to the following recipes: 

$\bf{Recipe \ A}$: First-order patterns generated  with $J = 1$, $\mu \in \{0, 0.1, \ldots, 0.9\}$, and $\theta_s \in \{0^\circ, 60^\circ, \ldots,300^\circ\}$,   amounting to a total of 60 distinct patterns.  

$\bf{Recipe \ B}$: Patterns generated by a linear combination of up to $C=4$ random \ac{DMA} patterns, given by
\vspace*{-0.15cm}
\begin{equation}\label{eqn:linearCombi}
\Lambda_{t}(\theta) = \frac{\sum_{c=1}^{C}\Lambda_{t, c}(\theta)}{\max\limits_{i} \left[\sum_{c=1}^{C} \Lambda_{t,c}(\theta)\right]_{i}}, \quad \forall t, 
\end{equation}
where each random DMA pattern $\Lambda_{t, c}(\theta)$ is generated using \eqref{eqn:dma_new} with a random $\mu$ from $\{0, 0.1, \ldots, 0.9\}$, a random angle $\theta_s$ from $\theta_s \in [0, 2\pi]$, and a random integer $J$ from $[1, 11]$.

$\bf{Recipe \ B+}$: An extended version of Recipe~B, which contains patterns from the DMA and rectangular patterns. In particular, 33.3\% of patterns are generated using Recipe~B, 33.3\% of the patterns are linear combinations of random rectangular patterns (random $\theta_{\mathrm{start}}$ and $\theta_{\mathrm{end}}$), and the remaining patterns are linear combinations of random DMA and rectangular patterns.

\section{Experimental Setup}
\subsection{Datasets and configurations}
We used speech signals from the `train-clean-360' and `dev-clean' subsets of the LibriSpeech corpus \cite{librispeech} as the source signals for training and validation, respectively. For the test sets, speech samples were selected from the EARS dataset \cite{richter2024ears} with minimum loudness of $-42$~dBFS \cite{loudness} as the selection criterion. The number of source-array setups for training and validation was $4320$ and $1080$, and each source-array setup corresponds to $P = 60$ user-defined patterns, resulting in a total of $4320 \times 60$ and $1080 \times 60$ samples for the training set and validation set, respectively. The number of test samples was $3240$ for each pattern under test. The duration per sample was 4~\unit{\s} for all training datasets. We set up to three concurrent sources per training sample, while fixing two concurrent sources per test sample \cite{ndf_iwaenc, steerable_NDF_FA_2025}. The source \acp{DOA} for training and validation are selected from $\theta_{n} \in \{0^{\circ}, 5^{\circ}, \ldots, 355^{\circ} \}$ and $\theta_{n}  \in \{2.5^{\circ}, 7.5^{\circ}, \ldots, 357.5^{\circ} \}$ respectively. For testing, $\theta_{n} \in \{1.25^{\circ}, 3.75^{\circ}, \ldots, 358.75^{\circ} \}$. The source-array distance $d = 1.5$~\unit{\m}. We set $L=72$ for the pattern vectors. All models were trained for $100$~epochs with a batch size of $10$ and an initial learning rate of $0.001$. Maximum suppression for all patterns was set to $-20$~\unit{\decibel}. $\epsilon$ in loss function \eqref{eqn:loss_func} was set to $ 10^{-7}$. We employed a four-microphone array ($Q=4$) with a diameter of \qty{3}{\cm}, consisting of three microphones arranged in a \ac{UCA} and an additional microphone at the array's center. For the remaining settings, we used the configurations in \cite{steerable_NDF_FA_2025}. 

\subsection{Performance measures}
\noindent\textbf{Directivity pattern}: For each test sample, we apply the estimated mask $\mathcal{M}[f, t]$ separately to each source at the reference microphone. Subsequently, a wideband power ratio $\xi[ \theta_n]$ for the $n$-th source is then calculated as
\vspace*{-0.2cm}
\begin{equation}\label{eqn:patternEstimation}
\xi[ \theta_n] = \frac{ \sum_{f=1}^{F}\sum_{t=1}^{T} \left| \mathcal{M}[f, t] \; X_{1, n}[f, t] \right|^2}{\sum_{f=1}^{F}\sum_{t=1}^{T}\left| X_{1, n}[f, t] \right|^2}.
\end{equation}
We then compute the arithmetic mean of the ratios ($\xi[ \theta_n]$) over all the test samples from the same direction $\theta_n$ to obtain the final estimated wideband directivity pattern. A narrowband directivity pattern is obtained by removing the summations over the frequency index $f$ in \eqref{eqn:patternEstimation}.

\noindent\textbf{SDR}: We use the \ac{SDR} \cite{vincent2006performance,von2022sa}, averaged over the test set, to measure the distortion in the estimated signals compared to the target signals.

\section{Results and Analysis}

To the best of our knowledge, parametric filtering \cite{thiergart2014informed} and the \ac{LS} beamformer \cite{ls_beamforming} remain the only effective methods for spatial filtering with configurable directivity patterns. Since parametric filtering significantly outperforms the \ac{LS} beamformer \cite{ndf_iwaenc}, we use it as a baseline for comparison. The parametric filter is computed using oracle DOA estimates, providing an upper bound on its performance.

\subsection{Conditioning methods}
\begin{figure}[t]
    \centering
	\begin{minipage}[b]{0.431\linewidth}
		\centering
		\centerline{ \includegraphics[width=\linewidth]{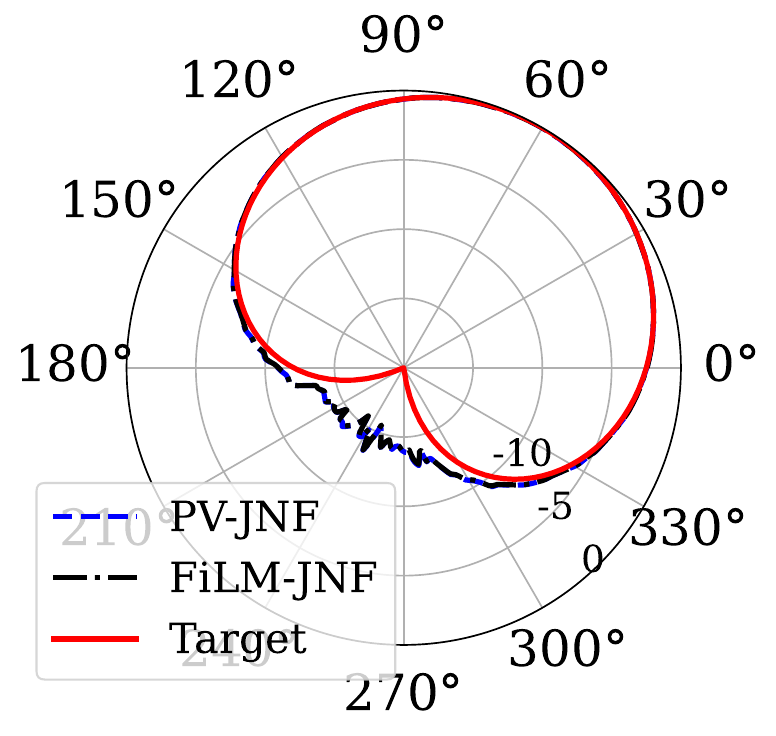}}
		(a)  $1^{\mathrm{st}}$-order pattern ($60^\circ$)  \\
	\end{minipage} 
            \begin{minipage}[b]{0.431\linewidth}
		\centering
		\centerline{ \includegraphics[width=\linewidth]{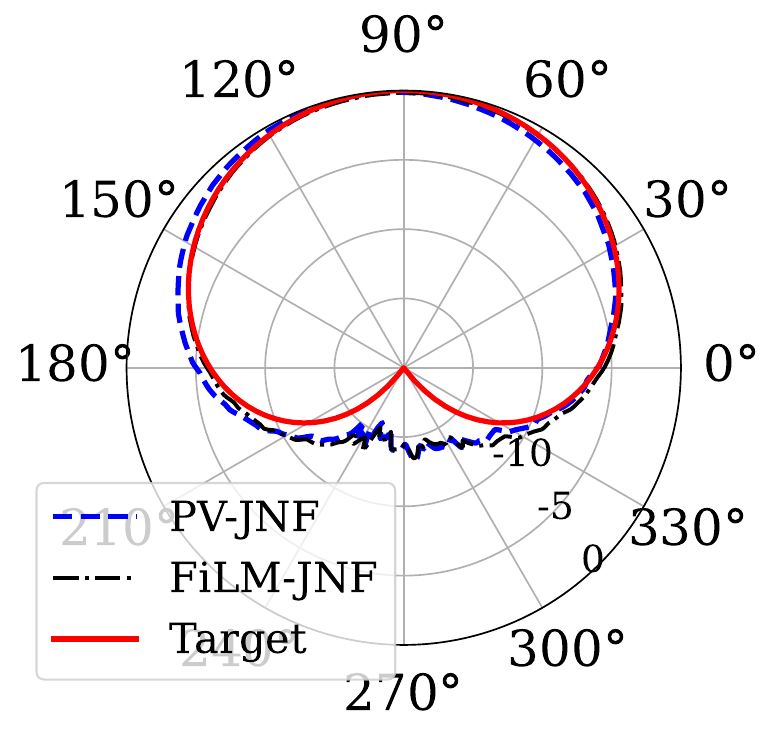}}
		(b) $1^{\mathrm{st}}$-order pattern ($90^\circ$)
	\end{minipage}
        \begin{minipage}[b]{0.431\linewidth}
		\centering
		\centerline{ \includegraphics[width=\linewidth]{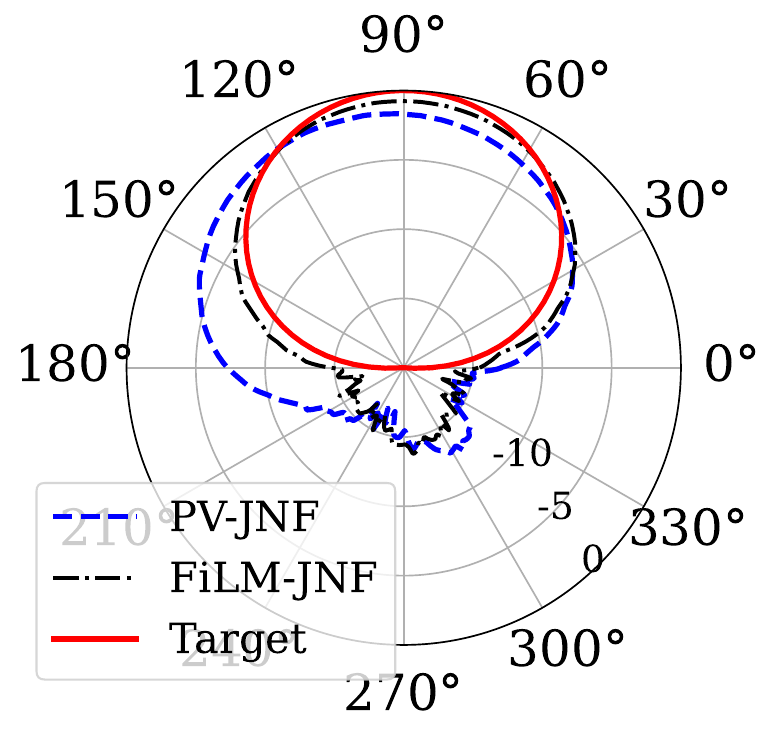}}
		(c) $3^{\mathrm{rd}}$-order pattern ($90^\circ$)
	\end{minipage}
            \begin{minipage}[b]{0.431\linewidth}
		\centering
		\centerline{ \includegraphics[width=\linewidth]{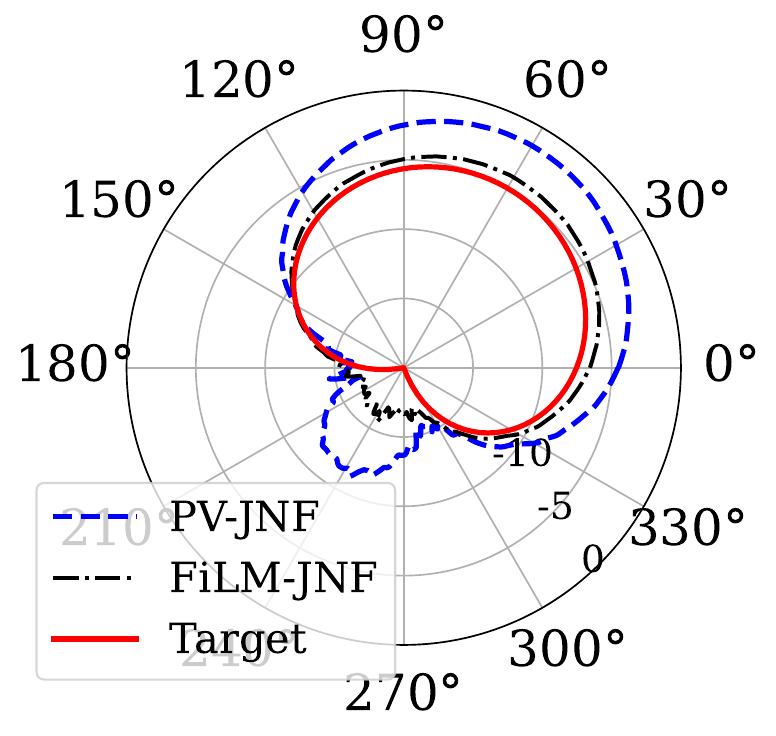}}
		(d) Scaled pattern 
	\end{minipage}        
    
      \caption{Estimated wideband directivity patterns obtained using all samples in the test set comparing PV-JNF and FiLM-JNF. Conditions: \ac{UNDF} models are trained using the training strategy `Recipe~A'. The scaled pattern is $-5$~\unit{\decibel} scaled version of the $1^{\mathrm{st}}$-order pattern. }
	\label{fig: film}  
        \vspace{-0.5em}
\end{figure}

\begin{table}[t]
  \centering
  \vspace{-6pt}
  \caption{SDR (\si{\decibel}) performance: comparison of proposed method with conditioning variants and baseline.}
  \label{tab: sdr_film}
  \resizebox{.482\textwidth}{!}{%
    \begin{tabular}{l rrrr}
    \toprule
       \multicolumn{1}{c}{} &\multicolumn{1}{c}{$1^{\mathrm{st}}$-order ($60^\circ$)}&\multicolumn{1}{c}{$1^{\mathrm{st}}$-order} ($90^\circ$)&\multicolumn{1}{c}{$3^{\mathrm{rd}}$-order ($90^\circ$)}&\multicolumn{1}{c}{Scaled} \\
       \midrule
       Parametric filtering \cite{thiergart2014informed} & 20.37  & 20.34  & 20.37 & 21.35\\
       PV-JNF (Recipe~A) & 26.50  & 20.87 & 10.40 & 8.11 \\
	    FiLM-JNF (Recipe~A)  & 26.30 & 24.31  & 17.34 & 19.54 \\   
	FiLM-JNF (Recipe~B)  & \bf{26.83} & \bf{26.30} &  \bf{24.54} & \bf{26.80} \\        
      \bottomrule
    \end{tabular}%
  }
  \vspace{-1em}
\end{table}


We study and compare the two conditioning methods in FiLM-JNF and PV-JNF using Recipe~A, which consists only of a finite number of specific $1^{\mathrm{st}}$-order patterns. Figure~\ref{fig: film} (a) shows the pattern estimates for a $1^{\mathrm{st}}$-order cardioid pattern steered towards $\theta_s = 60^\circ$, which is seen in the training. The estimated patterns of the two methods closely follow the target, except for limited attenuation towards the null, which was also observed in \cite{ndf_iwaenc}. Figure~\ref{fig: film} (b) shows the pattern estimates for $\theta_s = 90^\circ$ (unseen $\theta_s$ during training). We observe a slight steering bias for the PV-JNF while FiLM-JNF maintains a good approximation. Figure~\ref{fig: film} (c) illustrates the generalization performance for a $3^{\mathrm{rd}}$-order cardioid pattern (unseen order and $\theta_s$ during training). We observe that PV-JNF exhibits a larger steering error and poorer pattern approximation than FiLM-JNF, indicating the superiority of FiLM-based conditioning. Figure~\ref{fig: film} (d) shows the estimates for a scaled pattern, where the gain in the target direction is not unity. We see that PV-JNF deviates significantly from the target, whereas FiLM-JNF does not. The \ac{SDR} results in Table~\ref{tab: sdr_film} reflect the above observations, where the FiLM-JNF has significantly better performance (up to 11~\unit{\decibel}) than PV-JNF for the unseen scenarios. Thus, we conclude that FiLM conditioning enables the \ac{UNDF} to generalize to unseen higher-order patterns, unseen scaling, and unseen steering directions. In the following, we report the results for the \ac{UNDF} models using the FiLM-JNF architecture.

\subsection{Training strategies}

\begin{figure}[t!]
    \centering
	\begin{minipage}[b]{0.431\linewidth}
		\centering
		\centerline{ \includegraphics[width=\linewidth]{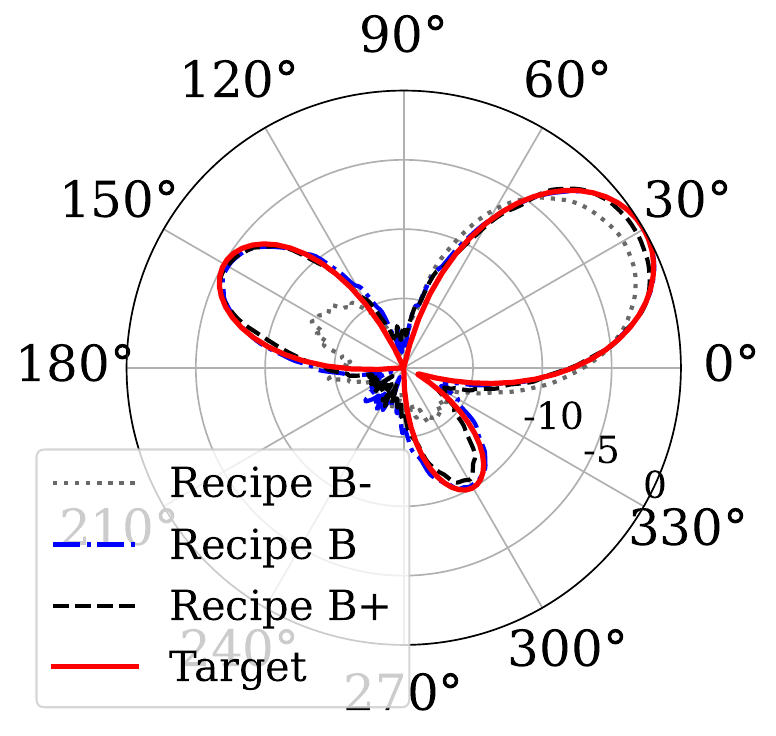}}
		(a)  Multi-scaled pattern  \\
	\end{minipage} 
            \begin{minipage}[b]{0.431\linewidth}
		\centering
		\centerline{ \includegraphics[width=\linewidth]{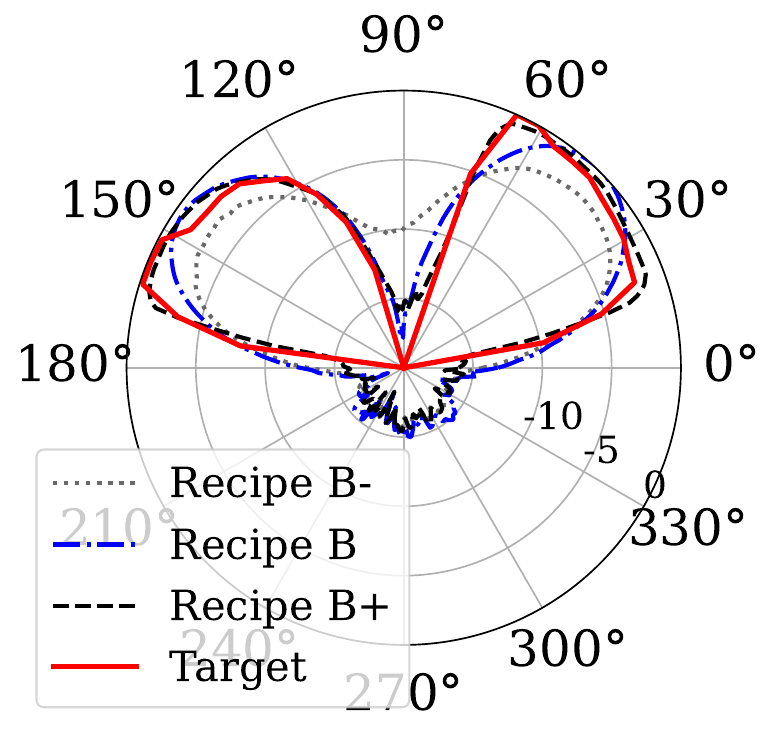}}
		(b) Irregular pattern 
	\end{minipage}
       
      \caption{Estimated wideband directivity patterns for the models using training strategies: Recipe~B-, Recipe~B, and Recipe~B+. Conditions: NDF models are trained using FiLM-JNF. }
	\label{fig:Film_recipe_w}  
      \vspace{-1em}
\end{figure}

\begin{figure}[t!]
    \centering
	\begin{minipage}[b]{0.4\linewidth}
		\centering
		\centerline{ \includegraphics[width=\linewidth]{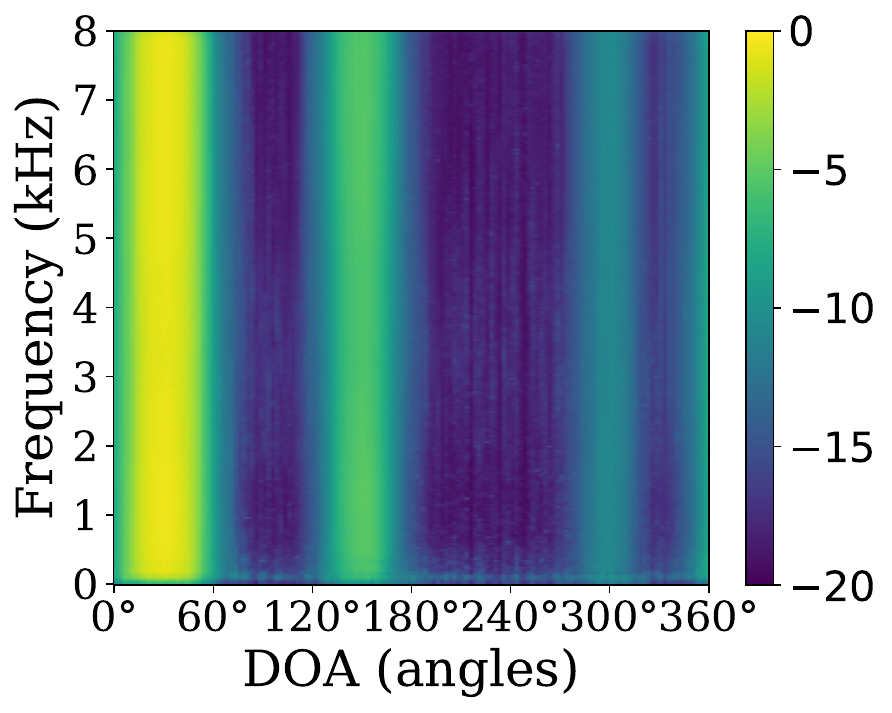}}
		(a)  Multi-scaled pattern  \\
	\end{minipage} 
            \begin{minipage}[b]{0.4\linewidth}
		\centering
		\centerline{ \includegraphics[width=\linewidth]{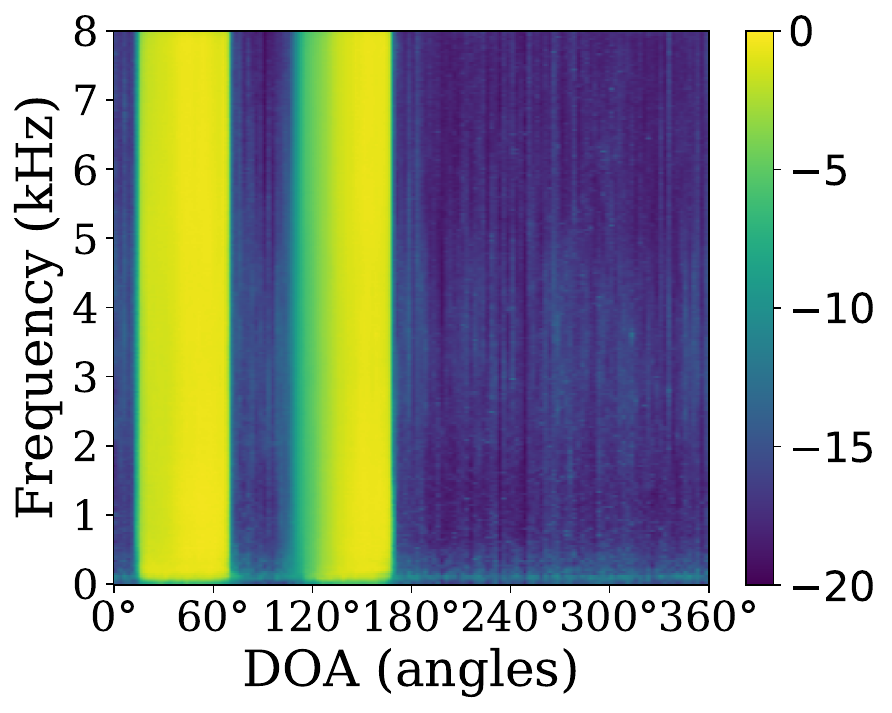}}
		(b) Irregular pattern 
	\end{minipage}
      \caption{Estimated narrowband directivity patterns for the models based on the FiLM-JNF trained by strategy Recipe~B+.}
	\label{fig:Film_recipe_n}  
    \vspace{-1em}
\end{figure}

\begin{table}[t!]
  \centering
  \vspace{-6pt}
  \caption{SDR (\si{\decibel}) performance on different training strategies.}
  \label{tab:sdr_recipe}
  \resizebox{.35\textwidth}{!}{%
    \begin{tabular}{l rrr}
    \toprule
       \multicolumn{1}{c}{} &\multicolumn{1}{c}{Recipe~B-}&\multicolumn{1}{c}{Recipe~B}&\multicolumn{1}{c}{ Recipe~B+} \\
       \midrule
       Multi-scaled pattern & 12.98  & \bf{19.75}  & 17.61  \\
	   Irregular pattern & 14.20 & 14.81  & \bf{20.39}  \\   
    \bottomrule
    \end{tabular}%
  }
  \vspace{-1.0em}
\end{table}

We trained the FiLM-JNF using Recipe~B and compared it with the parametric filter. As shown in Table~\ref{tab: sdr_film}, Recipe~B outperforms Recipe~A and the parametric filtering. Recipe~B differs from Recipe~A in two main ways. First, Recipe~B uses random \ac{DMA} patterns, which intuitively provide broader coverage of orders and steering directions during training. Second, Recipe~B involves linear combinations of DMA patterns. To better understand the impact of these combinations, we created a variant called Recipe~B-, which contains only one random \ac{DMA} pattern, i.e., with $C=1$ in \eqref{eqn:linearCombi}.  

We compare Recipes B-, B, and B+ based on pattern approximation in Figure~\ref{fig:Film_recipe_w}. For the multi-scaled pattern in Figure~\ref{fig:Film_recipe_w}, Recipe~B shows substantial improvement over Recipe~\mbox{B-, while} Recipe~B+ performs comparably to Recipe~B. We attribute this to the linear combination operation in Recipes B and B+, which exposes the model to a broader range of pattern variations. For the irregular pattern in Figure~\ref{fig:Film_recipe_w} (b), Recipe~B+ achieves the best approximation, demonstrating that rectangular pattern mixing in Recipe~B+ further enhances the approximation of irregular shapes. The \acp{SDR} in Table~\ref{tab:sdr_recipe} aligns with the observation in the pattern approximation. Furthermore, Figure~\ref{fig:Film_recipe_n} shows the estimated narrowband directivity patterns for strategy Recipe~B+, where the estimated patterns exhibit frequency invariance.

\subsection{Application for time-varying directivity patterns}
\begin{figure}[t!]
    \centering
	\begin{minipage}[b]{0.8\linewidth}
		\centering
		\centerline{ \includegraphics[width=\linewidth]{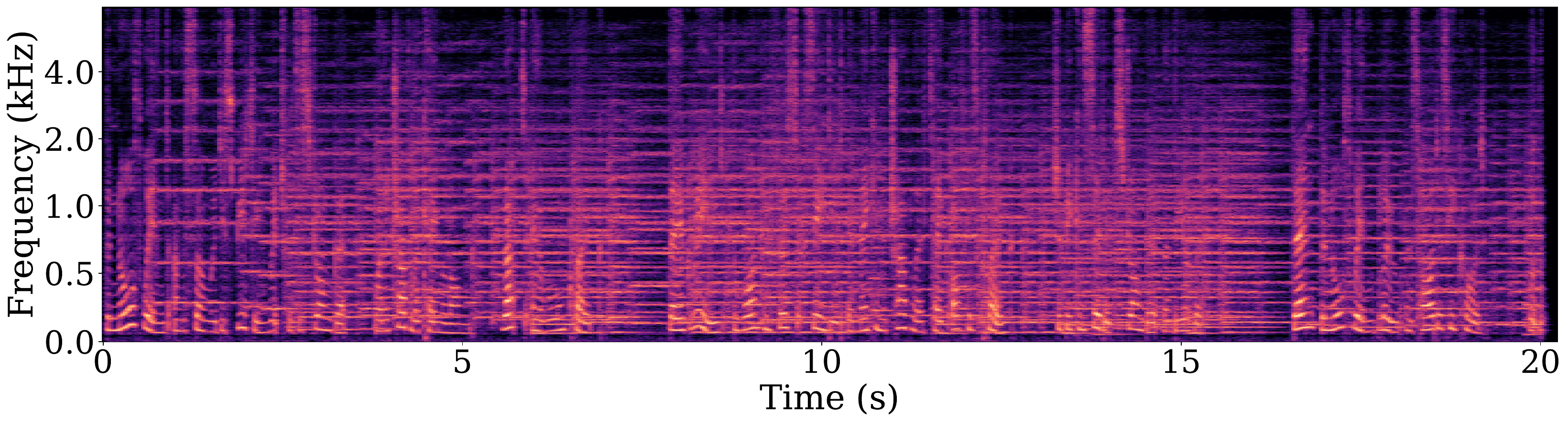}}
		(a)  Unprocessed spectrogram  \\ \vspace{0.5em}
	\end{minipage} 
            \begin{minipage}[b]{0.8\linewidth}
		\centering
		\centerline{ \includegraphics[width=\linewidth]{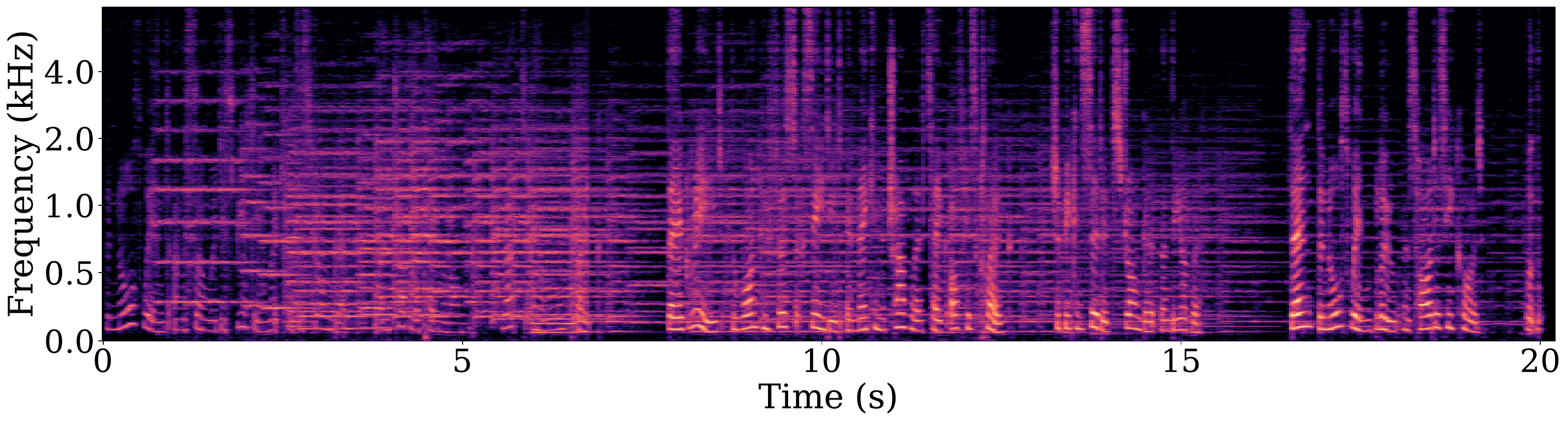}}
		(b) Processed spectrogram \\ \vspace{0.5em}
	\end{minipage}
            \begin{minipage}[b]{0.8\linewidth}
		\centering
		\centerline{ \includegraphics[width=\linewidth]{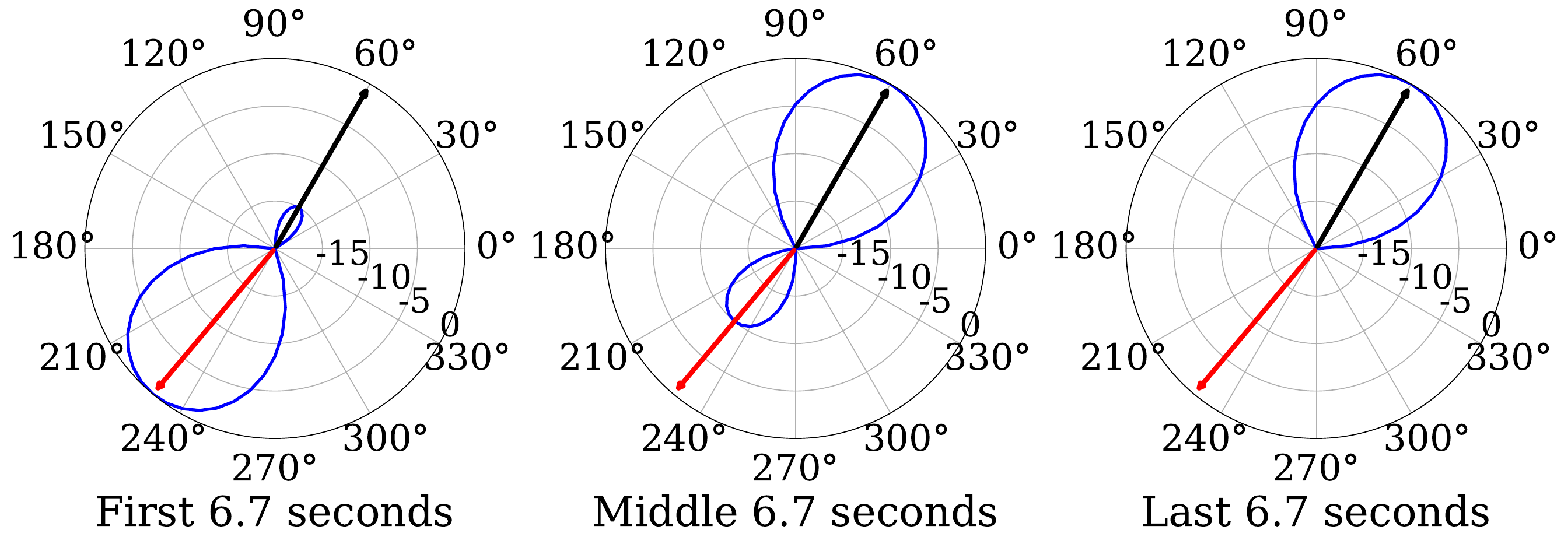}}
		(c) Pattern inputs for three distinct periods.   \\
	\end{minipage}       
      \caption{Applications for dynamic directivity patterns input.}
              \vspace{-1em}
	\label{fig:applications}  
\end{figure}

To illustrate the application of a time-varying directivity pattern, we considered a sound scene of duration $20$~\unit{\s} involving two concurrent sound sources in a reverberant room, a speech source located at an angle of $60^\circ$ and a music source located at an angle of $230^\circ$, both at a distance of $d = 1.5$~\unit{\m} from the array center. The simulated room \cite{RIRGenerator} was 6~\unit{\m} x 4~\unit{\m} x 3.5~\unit{\m} and had an $\mathrm{RT}_{60}$ of 0.15~\unit{\s}. Audio samples can be found online\footnote{\url{https://www.audiolabs-erlangen.de/resources/2026-EUSIPCO-UNDF}}. The spectrogram of the unprocessed mixture signal at the reference microphone is shown in Figure~\ref{fig:applications}(a). The entire simulation was divided into three equal time periods. The patterns shown in Figure~\ref{fig:applications} (c) are used as conditioning inputs for each of these consecutive periods. This process yields the spectrogram of the \ac{UNDF} output in Figure~\ref{fig:applications}(b). We used the \ac{UNDF} model, trained with the FiLM-JNF architecture using Recipe~B+ for this illustration. In the first $6.7$~\unit{\s}, the music components in the mixture are preserved, and the speech components are suppressed to a certain level. In the middle $6.7$~\unit{\s}, the speech components recover immediately, and the music components are partly suppressed. In the last $6.7$~\unit{\s}, the retained music components are suppressed further. These results are in agreement with the expected results given the patterns shown in Figure~\ref{fig:applications}(c). Hence, this qualitative example illustrates that users can adjust the proportions of each component in the processed mixture by varying the input directivity pattern. Because conditioning is applied to every frame and the model is causal \cite{tesch_insights,ndf_iwaenc}, changes in the input pattern are reflected immediately in the output in our simulation. This experiment also demonstrates that the \ac{UNDF} model, despite being trained solely on anechoic speech signals, generalizes to non-speech signals and low-reverberant acoustic conditions.

\section{Conclusions}
\label{sec:con}
We propose UNDF, a spatial filtering method with configurable directivity patterns at inference. To achieve this, we integrate the FiLM layer into the JNF architecture, enabling conditioning on user-defined patterns. We demonstrated that the proposed FiLM-JNF architecture enables UNDF to generalize to unseen higher orders, scaling variations, and steering directions. Progressive evolution of training strategies improves pattern approximation, enabling the approximation of irregular shapes.

\section{Acknowledgment}
The authors gratefully acknowledge the scientific support and HPC resources provided by the Erlangen National High Performance Computing Center (NHR@FAU) of the Friedrich-Alexander-Universität Erlangen-Nürnberg (FAU). The hardware is funded by the German Research Foundation (DFG).

\bibliographystyle{IEEEbib}
\bibliography{refs}

\end{document}